\documentclass[letter,twocolumn]{jpsj2} %% for letters
\usepackage{graphicx}
\usepackage{amssymb}
\title{Successive Phase Transitions in Na$_{0.5}$CoO$_{2}$ based on the Triangular Lattice d-p Model}
\author{Youichi \textsc{Yamakawa$^1$}\thanks{E-mail address: yamakawa@phys.sc.niigata-u.ac.jp}, 
and Yoshiaki \textsc{\={O}no}$^{1,2}$}
\inst{$^1$Department of Physics, Niigata University, Ikarashi, Nishi-ku, Niigata 950-2181, Japan

$^2$Center for Transdisciplinary Research, Niigata University, Ikarashi, Nishi-ku, Niigata 950-2181, Japan}

\abst{
We investigate the electronic states of the CoO$_2$ plane in the layered cobalt oxides Na$_{0.5}$CoO$_2$ using the realistic 11 band d-p model on a two-dimensional triangular lattice. Effects of the Coulomb interaction on a Co site: the intra- and inter-orbital direct terms $U$ and $U'$, the exchange coupling $J$ and the pair-transfer $J'$, are treated within the Hartree-Fock approximation. 
It is found that the metallic antiferromagnetism takes place below $T_{c1}$ and, in addition, the orbital order accompanied by the small charge order takes place below $T_{c2} (< T_{c1})$ where the system becomes insulator. The obtained results are consistent with the successive phase transitions observed in Na$_{0.5}$CoO$_{2}$. 
}

\kword{
cobalt oxides, antiferromagnetism, charge order, orbital order, metal-insulator transition}
\begin{document}
\maketitle

%%%%%%%%%%%%%%%%%%%%%%%%%%%%%%%%%%%%%%%%%%%%%%%%%%%%%%%%%%%%%%%%%%%%%%%%%%%%
%\section{Introduction}
The discovery of superconductivity in the layered cobalt oxide Na$_{x}$CoO$_{2} \cdot$yH$_{2}$O \cite{Takada} has stimulated further interest in the electronic states of the mother compound Na$_x$CoO$_2$. 
The specific features of the system are the geometrical fluctuation of the CoO$_2$ plane which consists of a two dimensional triangular lattice of Co atoms, and the orbital degeneracy of three $t_{2g}$ bands for Co $3d$ electrons. 
The electron filling in the CoO$_2$ plane is controlled by changing the Na content $x$: the  electron filling of the  three $t_{2g}$ bands is given by $5+x$. 
For $x < 0.6$, Na$_x$CoO$_2$ is a normal paramagnetic metal except for $x = 0.5$, while, for $0.6 < x < 0.75$, an anomalous behavior is observed: the magnetic susceptibility $\chi$ is Curie-Weiss like although the resistivity $\rho$ is metallic \cite{Foo}, the electronic specific heat coefficient $\gamma$ is large and increases with $x$ \cite{Yokoi}. For $x > 0.7$, Na$_x$CoO$_2$ exhibits large thermopower  and is a good prospect for thermoelectric applications \cite{Terasaki, Lee, Koshibae}. 
For $x > 0.75$, a weak magnetic order is observed below 20 K \cite{Motohashi}, where the ferromagnetic ordered CoO$_2$ layers couple antiferromagnetically with each other \cite{Bayrakci}. 

Furthermore, Na$_{0.5}$CoO$_2$ exhibits remarkable successive phase transitions at $T_{c1}\sim 87$ K and  $T_{c2}\sim 53$ K \cite{Foo}: the itinerant antiferromagnetism is realized below $T_{c1}$ and the system becomes insulator below $T_{c2}$. 
The magnetic susceptibility $\chi$ shows kinks at $T_{c1}$ and $T_{c2}$. 
The resistivity $\rho$ exhibits only a tiny anomaly at $T_{c1}$, while it rapidly increases below $T_{c2}$. From the NMR and the neutron measurements, Yokoi {\it et  al.} \cite{Yokoi} have proposed the magnetic structure of Na$_{0.5}$CoO$_2$ below $T_{c1}$, where chains of Co$^{3.5+\delta}$ with larger staggered moments within the CoO$_2$ ($ab$) plane and chains of Co$^{3.5-\delta}$ with smaller moments along the $c$ axis align alternatively on the $ab$ plane. 
The ordering of the $c$ axis oriented moments on the Co$^{3.5-\delta}$ chains has not been distinguished whether ferromagnetic or antiferromagnetic \cite{Yokoi}, and has not been detected by polarized neutron measurements \cite{Gasparovic}. 
The charge disproportionation of the Co sites into chains of Co$^{3.5+\delta}$ and Co$^{3.5-\delta}$ is closely related to the ordered pattern of Na ions which form onedimensional zigzag chains below room temperature \cite{Huang, Znadbergen}. 

From the zero-field NMR spectra which exhibit splittings below $T_{c2}$ \cite{Yokoi}, Ning {\it et al.} \cite{Ning} have proposed the charge ordering pattern below $T_{c2}$, where the Co valence alternates on the Co$^{3.5-\delta}$ chains while it is constant on the Co$^{3.5+\delta}$ chains. 
The zero-field NMR measurements have also been carried out for K$_{0.5}$CoO$_2$ \cite{Watanabe,Yokoi2}, which exhibits similar successive transitions at $T_{c1}\sim 60$ K and  $T_{c2}\sim 20$ K, but the splitting below $T_{c2}$ has not been observed in contrast to Na$_{0.5}$CoO$_2$. 
Then, the origin of the metal-insulator transition at $T_{c2}$ still remains controversial. 

%%%%%%%%%%%%%%%%%%%%%%%%%%%%%%%%%%%%%%%%%%%%%%%%%%%%%%%%%%%%%%%%%%%%%%%%%%%%
\begin{figure}[b]
\begin{center}
\includegraphics[scale=0.275]{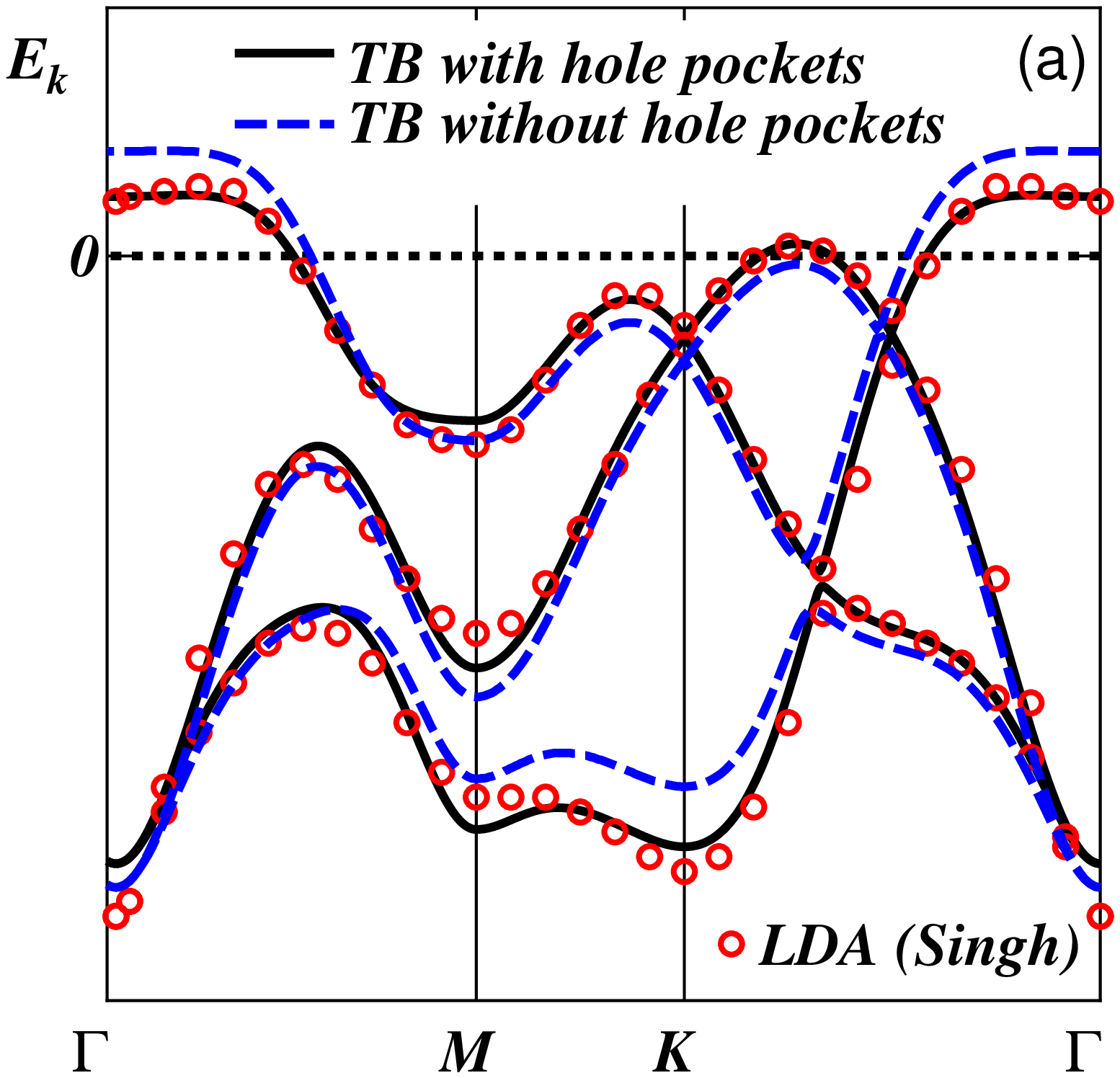}
\includegraphics[scale=0.20]{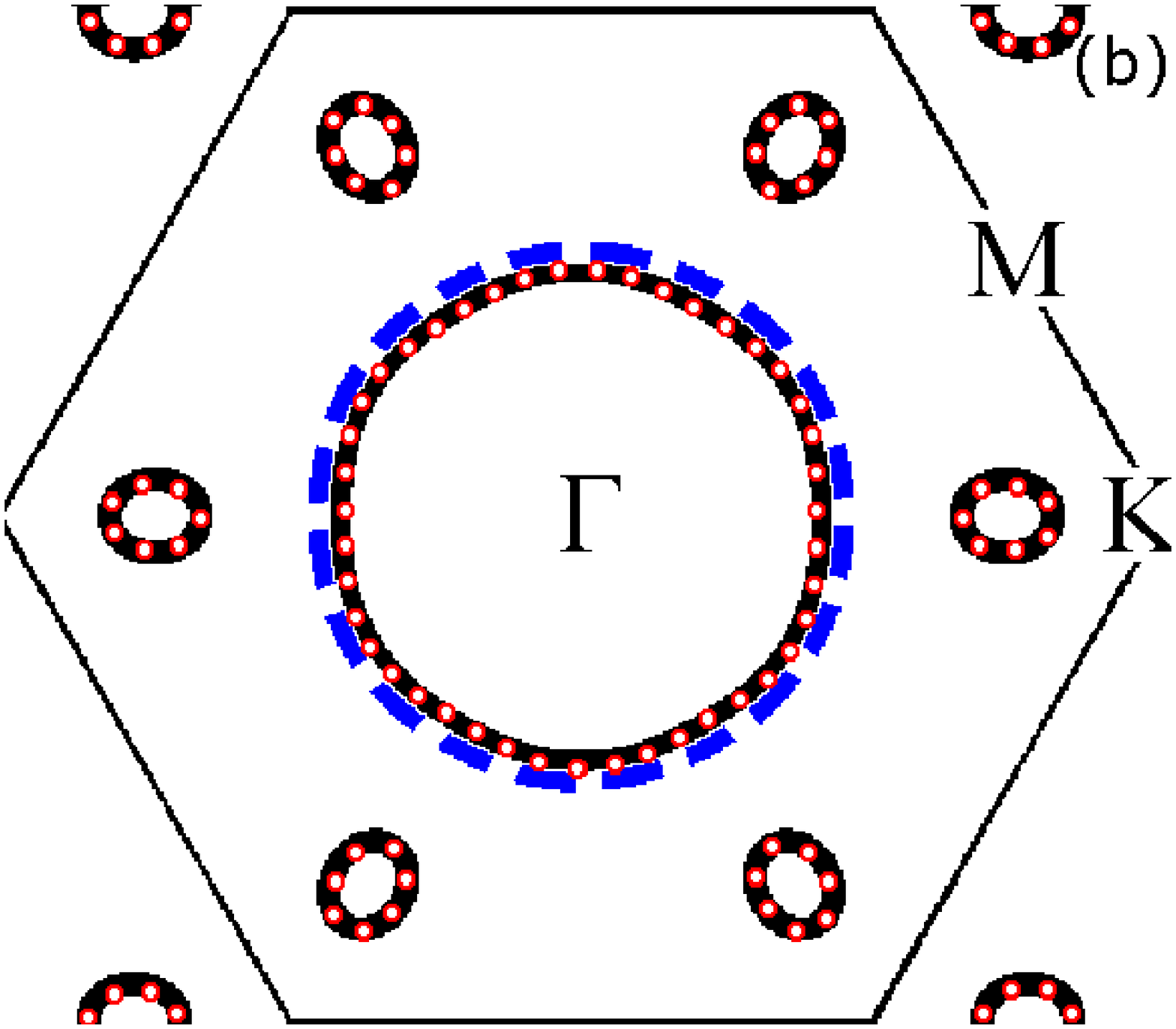}
\caption{
The band structure (a) and the Fermi surface (b) from the $d$-$p$ tight-binding model with hole pockets (solid lines), without hole pockets (dashed lines) and from the LDA \cite{Singh} (open circles), 
}
\label{fig-lda}
\end{center}
\end{figure}
%%%%%%%%%%%%%%%%%%%%%%%%%%%%%%%%%%%%%%%%%%%%%%%%%%%%%%%%%%%%%%%%%%%%%%%%%%%%

Many theoretical works have been done on the electronic states of Na$_{0.5}$CoO$_2$ for the last few years \cite{Li,Indergand,Bobroff,Choy,Zhou}. Some of the proposed magnetic structures are consistent with the experimental prediction below $T_{c1}$ \cite{Yokoi,Gasparovic}, however, none of the proposed charge ordering patterns are consistent with the experimental prediction below $T_{c2}$ \cite{Ning}. 
In our previous works \cite{Yamakawa, Yamakawa2}, the electronic states of Na$_{0.5}$CoO$_2$ have been investigated  using the 11 band d-p model on a two-dimensional triangular lattice, where the tight-binding parameters are determined so as to fit the LDA band structure \cite{Singh} (see Fig. \ref{fig-lda}). 
Effects of the Coulomb interaction at a Co site are treated within the Hartree-Fock approximation, and the effect of the onedimensional Na order is considered by taking into account of an effective potential on the CoO$_2$ plane. It was found that the Na order enhances the Fermi surface nesting, which is suppressed due to the frustration effect in the case without the Na order, and results in the metallic antiferromagnetism \cite{Yamakawa}, whose magnetic structure is consistent with the experimental prediction \cite{Yokoi,Gasparovic}.

In addiction, we have considered the effect of the Coulomb interaction between the nearest-neighbor Co sites $V$ and found that a coexistence state of the antiferromagnetic, charge and orbital order takes place for $V>V_c$, where the system becomes insulator \cite{Yamakawa2}. The obtained charge ordering pattern is consistent with the experimental prediction by Ning {\it et al.} \cite{Ning}, but the charge differentiation due to the charge order seems to be too large as compared to the  splittings of the zero-field NMR spectra, which are very small for Na$_{0.5}$CoO$_2$ and are absent for K$_{0.5}$CoO$_2$ \cite{Watanabe,Yokoi2}. 
The overestimation of the charge differentiation is considered to be due to the effect of the intersite interaction $V$ which is known to enhance the charge order.  Therefore, in the present work, we discuss the spin, orbital and charge ordered states exclusively due to the effects of the on-site interaction in the absence of the intersite interaction $V$.

%%%%%%%%%%%%%%%%%%%%%%%%%%%%%%%%%%%%%%%%%%%%%%%%%%%%%%%%%%%%%%%%%%%%%%%%%%%%
%\section{Model}
To investigate the electronic states of the CoO$_2$ plane in the layered cobalt oxides Na$_{x}$CoO$_2$, we employ the two dimensional triangular lattice $d$-$p$ model \cite{Yamakawa, Yamakawa2, Yada, Yada2} which includes 11 orbitals:  $d_{3z^2-r^2}$, $d_{x^2-y^2}$, $d_{xy}$, $d_{yz}$, $d_{zx}$ of Co and $p_{1x}$, $p_{1y}$, $p_{1z}$ ($p_{2x}$, $p_{2y}$, $p_{2z}$) of O in the upper (lower) side of the Co plane. 
The model is given by the following Hamiltonian: 
%%%%%%%%%%%%%%%%%%%%%%%%%%%%%%%%%%%%%%%%%%%%%%%%%%%%%%%%%%%%%%%%%%%%%%%%%%%%
\begin{eqnarray}
H  \hspace{-2mm}&=&\hspace{-2mm}
           \sum_{{\bf n},m,\sigma}
           \varepsilon_{{\bf n}m}^{d}
           d_{{\bf n}m\sigma}^{\dagger}
           d_{{\bf n}m\sigma}  
         + \sum_{{\bf k},m,m',\sigma}
           t_{{\bf k}mm'}^{dd}
           d_{{\bf k}m\sigma}^{\dagger}
           d_{{\bf k}m'\sigma} 
           \nonumber \\
   \hspace{-2mm}&+&\hspace{-2mm}
           \varepsilon_{p} \sum_{{\bf k},j,l,\sigma}
           p_{{\bf k}jl\sigma}^{\dagger}
           p_{{\bf k}jl\sigma} 
         + \sum_{{\bf k},j,j',l,l',\sigma}
           t_{{\bf k}jj'll'}^{pp}
           p_{{\bf k}jl\sigma}^{\dagger}
           p_{{\bf k}j'l'\sigma} 
           \nonumber \\
   \hspace{-2mm}&+&\hspace{-2mm}
           \sum_{{\bf k},j,l,m,\sigma}
          (t_{{\bf k}jlm}^{pd}
           p_{{\bf k}jl\sigma}^{\dagger}
           d_{{\bf k}m\sigma}
         + {\rm h.c.}) 
           \nonumber \\
   \hspace{-2mm}&+&\hspace{-2mm}
           U \sum_{{\bf n},m}
           \hat{n}^d_{{\bf n}m\uparrow}
           \hat{n}^d_{{\bf n}m\downarrow}
         + U'\sum_{{\bf n},m > m'}
           \hat{n}^d_{{\bf n}m}
           \hat{n}^d_{{\bf n}m'}
           \nonumber \\
   \hspace{-2mm}&+&\hspace{-2mm}
           J \sum_{{\bf n},m > m',\sigma,\sigma'}
           d_{{\bf n}m\sigma}^{\dagger}
           d_{{\bf n}m'\sigma'}^{\dagger}
           d_{{\bf n}m\sigma'}
           d_{{\bf n}m'\sigma} 
           \nonumber \\
   \hspace{-2mm}&+&\hspace{-2mm}
           J' \sum_{{\bf n},m > m',\sigma}
           d_{{\bf n}m\sigma}^{\dagger}
           d_{{\bf n}m-\sigma}^{\dagger}
           d_{{\bf n}m'-\sigma}
           d_{{\bf n}m'\sigma} ,
\label{eq-Model}
\end{eqnarray}
%%%%%%%%%%%%%%%%%%%%%%%%%%%%%%%%%%%%%%%%%%%%%%%%%%%%%%%%%%%%%%%%%%%%%%%%%%%%
where $d_{{\bf k}m\sigma}^{\dagger}$ ($d_{{\bf n}m\sigma}^{\dagger}$) is a creation operator for a Co $3d$ electron with wave vector ${\bf k} = (k_x, k_y)$ (site ${\bf n} = (n_x,n_y)$), orbital $m (= 3z^2-r^2, x^2-y^2, xy, yz, zx)$ and spin $\sigma (= \uparrow,  \downarrow)$,  and $p_{{\bf k}jl\sigma}^{\dagger}$ is a creation operator for a oxygen $2p$ electron with  wave vector ${\bf k}$, site $j (=1, 2)$, orbital $l (=x, y, z)$ and spin $\sigma$, respectively; 
$\hat{n}^d_{{\bf n}m\sigma}=d_{{\bf n}m\sigma}^{\dagger}d_{{\bf n}m\sigma}$ and 
$\hat{n}^d_{{\bf n}m}=\sum_\sigma \hat{n}^d_{{\bf n}m\sigma}$. 
The transfer integrals $t_{{\bf k},j,j',l,l'}^{pp}$, $t_{{\bf k},j,l,m}^{pd}$ and $t_{{\bf k},m,m'}^{dd}$, which are written by the Slater-Koster parameters, together with the atomic energies $\varepsilon_{p}$ and $\varepsilon_{m}^{d}$ are determined so as to fit the tight-binding energy bands to the LDA bands for Na$_{0.5}$CoO$_2$ \cite{Singh} as shown in Figs.\ref{fig-lda}(a) and (b). 
As the six hole pockets near the $K$ points predicted by the LDA \cite{Singh} have not been observed by ARPES experiments \cite{Yang}, we employ the energy bands without the hole pockets where the tight-binding parameters relating to the trigonal distortion are slightly modified (see Figs.\ref{fig-lda}(a) and (b)).

In the Hamiltonian eq. (\ref{eq-Model}), we consider the effects of the Coulomb interaction at a Co site: the intra- and inter-orbital direct terms $U$ and $U'$, the exchange coupling $J$ and the pair-transfer $J'$. 
%, and effects of the Inter-atomic Coulomb interaction between the nearest neighbor Co sites $V$. 
Here and hereafter, we assume the rotational symmetry yielding the relations: $U'=U-2J$ and $J=J'$, and we set $J = U/10$ for simplicity. 
We also consider the effect of the Na order at $x=0.5$, where Na ions form onedimensional chains below room temperature, by taking into account an effective onedimensional potential on the CoO$_2$ plane \cite{Yamakawa, Yamakawa2} : 
%%%%%%%%%%%%%%%%%%%%%%%%%%%%%%%%%%%%%%%%%%%%%%%%%%%%%%%%%%%%%%%%%%%%%%%%%%%%
\begin{eqnarray}
\label{eq-Na}
	\varepsilon_{{\bf n},m}^{d}
=	\left\{
	\begin{array}{ll}
		\varepsilon_{m}^{d} - \Delta \varepsilon_{d}
	&	\begin{array}{l}
			\qquad $for odd $n_y \\
			$(on Na ordered lines) $ \\
		\end{array} \\
		\varepsilon_{m}^{d} + \Delta \varepsilon_{d}
	&	\begin{array}{l}
			\qquad $for even $n_y\\
			$(out of Na ordered lines)$ \\
		\end{array}
	\end{array}
	\right. \!\!\!\!\!\!\!\!\!\!\!\!
\end{eqnarray}
%%%%%%%%%%%%%%%%%%%%%%%%%%%%%%%%%%%%%%%%%%%%%%%%%%%%%%%%%%%%%%%%%%%%%%%%%%%%
with the effective potential $\Delta \varepsilon_{d}$ due to the Na order. 
In this paper, we set $\Delta \varepsilon_{d} = 1.0$ eV. 
Due to the effect of $\Delta \varepsilon_{d}$, the band structure becomes quasi one dimensional and the Fermi surface nesting is enhanced resulting in the antiferromagnetic order\cite{Yamakawa, Yamakawa2}. 

%The onedimensional potential $\Delta \varepsilon_{d}$ that leads the antiferromagnetic ordered state \cite{Yamakawa} is set to 1.0 eV. 

%Hereafter, in order to consider the both Na order and the antiferromagnetic order, we use an extended unit cell including four Co atoms and eight O atoms, together with the magnetic Brillouin zone which is $\frac14$ of the original Brillouin zone. 

%%%%%%%%%%%%%%%%%%%%%%%%%%%%%%%%%%%%%%%%%%%%%%%%%%%%%%%%%%%%%%%%%%%%%%%%%%%%
%\section{Result}
%%%%%%%%%%%%%%%%%%%%%%%%%%%%%%%%%%%%%%%%%%%%%%%%%%%%%%%%%%%%%%%%%%%%%%%%%%%%
\begin{figure}[tb]
\begin{center}
\includegraphics[scale=0.4]{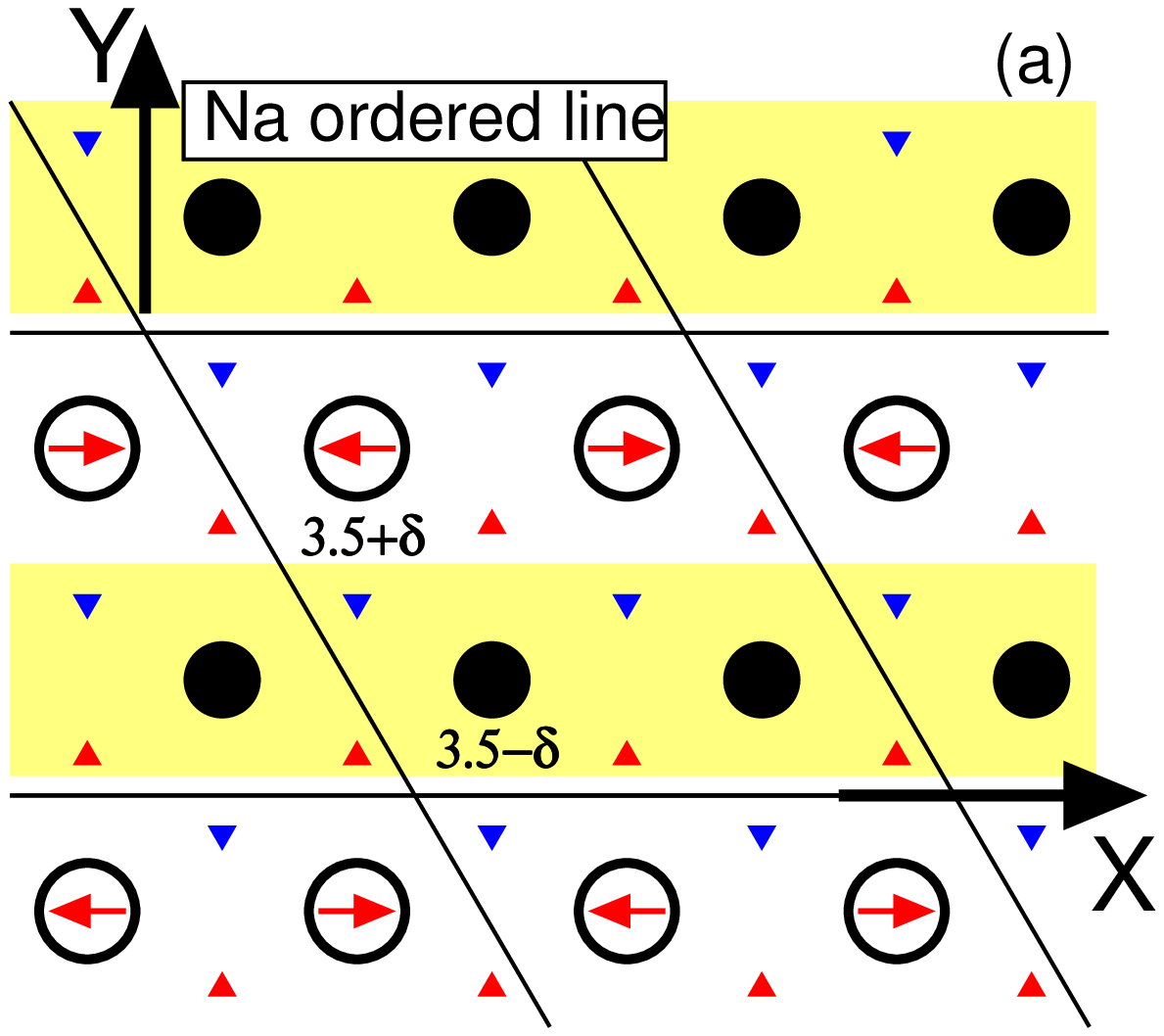}
\includegraphics[scale=0.4]{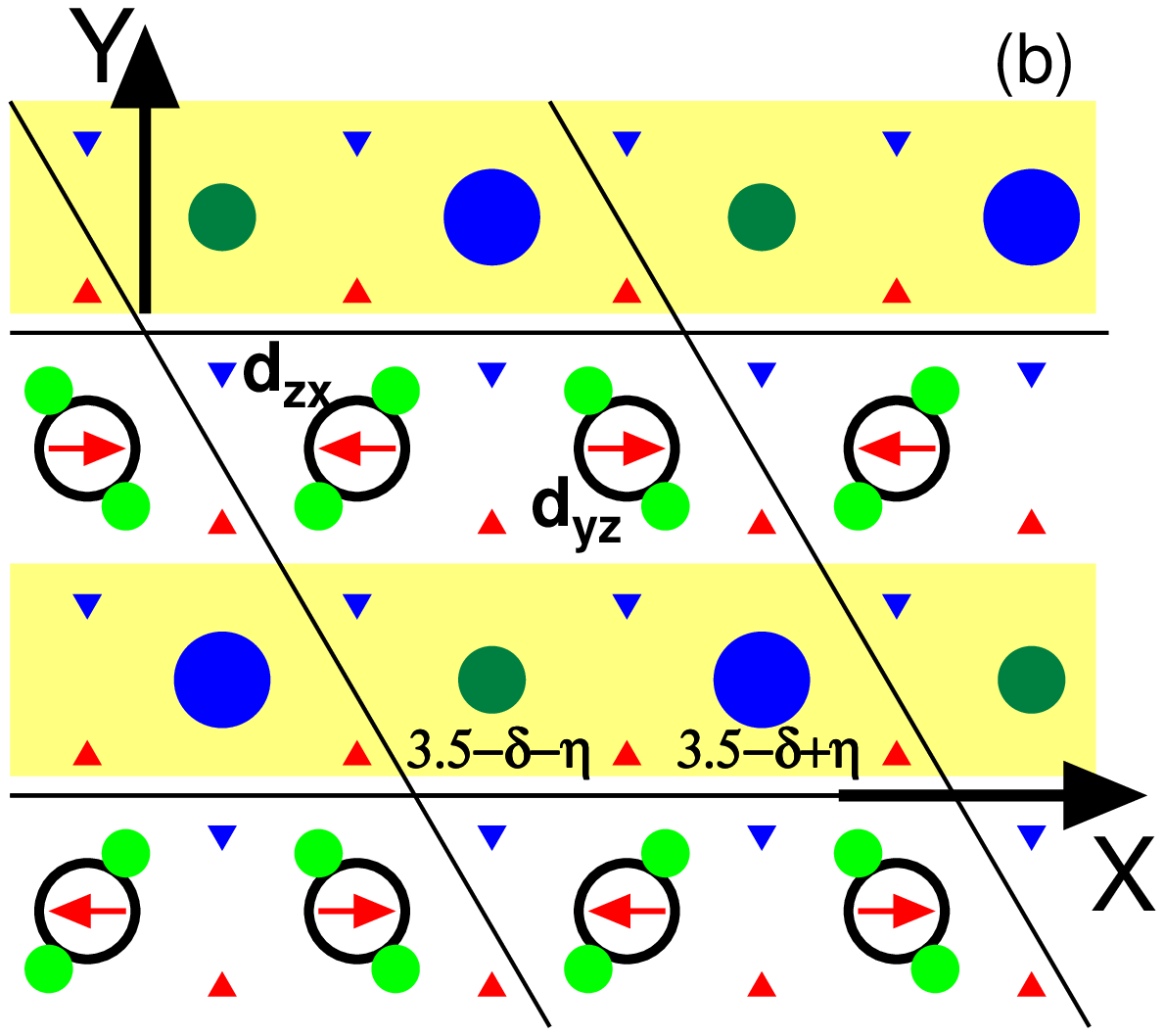}
\caption{
The schematic diagram of the antiferromagnetic ordered state (a) and the coexistence state of the antiferromagnetic, charge and orbital order (b). 
The filled (open) circles represent Co atoms on (out of) Na ordered lines and the upward (downward) triangles represent O atoms in the upper (lower) layers of the Co layer. The parallelogram is the extended unit cell. 
}
\label{fig-struct}
\end{center}
\end{figure}
%%%%%%%%%%%%%%%%%%%%%%%%%%%%%%%%%%%%%%%%%%%%%%%%%%%%%%%%%%%%%%%%%%%%%%%%%%%%

Now we discuss the possible ordered states of the Hamiltonian eq. (\ref{eq-Model}) within the Hartree-Fock approximation, where we assume that the order parameters are diagonal with respect to the orbital $m$ and the spin $\sigma$. 
Figure \ref{fig-struct}(a) shows the obtained ordered pattern of the antiferromagnetic state (AFM). Chains of Co$^{3.5+\delta}$ and Co$^{3.5-\delta}$ align alternately within the CoO$_2$ plane due to the effect of the onedimensional Na order \cite{Huang}. 
Figure \ref{fig-struct}(b) shows an order pattern of a coexistence state of antiferromagnetic, charge and orbital ordered state (AFM+CO+OO), where the Co sites out of the Na ordered line with the valency of 3.5+$\delta$ have the antiferromagnetic moment \cite{Yokoi, Gasparovic}, while, the Co sites on the Na ordered line with the valency of 3.5-$\delta$ show a charge order of Co$^{3.5+\delta+\eta}$ and Co$^{3.5+\delta-\eta}$ as proposed by Ning {\it et al.}\cite{Ning}. 
We note that the charge order in X-direction is different from the charge order of Co$^{3.5+\delta}$ and Co$^{3.5-\delta}$ in Y-direction due to the effect of the onedimensional Na order.

%\subsection{without holepocket}
%%%%%%%%%%%%%%%%%%%%%%%%%%%%%%%%%%%%%%%%%%%%%%%%%%%%%%%%%%%%%%%%%%%%%%%%%%%%
\begin{figure}[t]
\begin{center}
\includegraphics[scale=0.4]{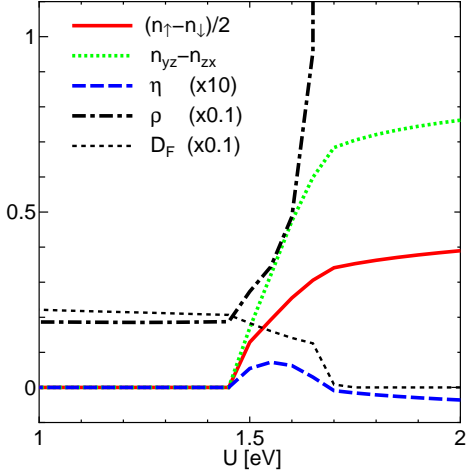}
\caption{
The sublattice magnetization $(n_{\uparrow}-n_{\downarrow})/2$, the orbital order $n_{yz}-n_{zx}$, the charge order $\eta$, the resistivity $\rho$ and the density of states at the Fermi level $D_{F}$ as functions of $U$ for $\Delta \varepsilon_d$ = 1.0 eV at $T$ = 20 K. 
}
\label{fig-udep}
\end{center}
\end{figure}
%%%%%%%%%%%%%%%%%%%%%%%%%%%%%%%%%%%%%%%%%%%%%%%%%%%%%%%%%%%%%%%%%%%%%%%%%%%%

Figure \ref{fig-udep} shows the sublattice magnetization $(n_{\uparrow}-n_{\downarrow})/2$, the orbital order $n_{yz}-n_{zx}$, the charge order $\eta$, the density of states at the Fermi level $D_{\rm F}$ and the resistivity $\rho$ as functions of $U$ at $T = 20$ K. 
%$\rho = 1/\left(\sigma_{xx}+\sigma_{yy}\right)$
Here we estimate the resistivity by using the Boltzmann equation as $\rho = 1/\left(\sigma_{xx}+\sigma_{yy}\right)$ with 
%%%%%%%%%%%%%%%%%%%%%%%%%%%%%%%%%%%%%%%%%%%%%%%%%%%%%%%%%%%%%%%%%%%%%%%%%%%%
\begin{equation}
\sigma_{\mu \mu}
=
\frac{e^2}{N\gamma}
\sum_{\vec{k} \sigma} 
\frac{\partial f}{\partial \varepsilon}
\left( 
\frac{1}{\hbar}
\frac{\partial \varepsilon}{\partial k_\mu}
\right)^2,
\qquad \left(\mu=x,y\right).
\end{equation}
%%%%%%%%%%%%%%%%%%%%%%%%%%%%%%%%%%%%%%%%%%%%%%%%%%%%%%%%%%%%%%%%%%%%%%%%%%%%
%The onedimensional potential $\Delta \varepsilon_{d}$ that leads the antiferromagnetic ordered state \cite{Yamakawa} is set to 1.0 eV. 
%Because the charge order is suppressed due to the effect of $U$, it is realistic though the $\Delta \varepsilon_{d}$ value is felt large. 
When $U$ is larger than a critical value $U_c \sim 1.45$ eV, we observe the coexistence state AFM+CO+OO schematically shown in Fig. \ref{fig-struct} (b). 
%The phase transition between the paramagnetic state (PM) and the coexistence state between antiferromagnetic, charge and orbital ordered state AFM+CO+OO takes place at $U_c = 1.45$ eV. 
When $U$ increases for $U > U_c$, both the sublattice magnetization $(n_{\uparrow}-n_{\downarrow})/2$ and the orbital order $n_{yz}-n_{zx}$ increase, while the density of states $D_F$ decreases and finally becomes zero for $U > 1.7$ eV. 
The resistivity $\rho$ rapidly increases for $U > U_c$ and becomes almost infinity for $U > 1.7$ eV. 
Due to the structure of the orbital order (see Fig. \ref{fig-struct} (b)), the charge order $\eta$ is induced by the orbital order although the amount of $\eta$ is vary small or zero as shown in Fig. \ref{fig-udep}. 

%As a result, the charge order $\eta$ is caused by the orbital order because of geometrical structure but it is very small. 

%%%%%%%%%%%%%%%%%%%%%%%%%%%%%%%%%%%%%%%%%%%%%%%%%%%%%%%%%%%%%%%%%%%%%%%%%%%%
\begin{figure}[t]
\begin{center}
\includegraphics[scale=0.45]{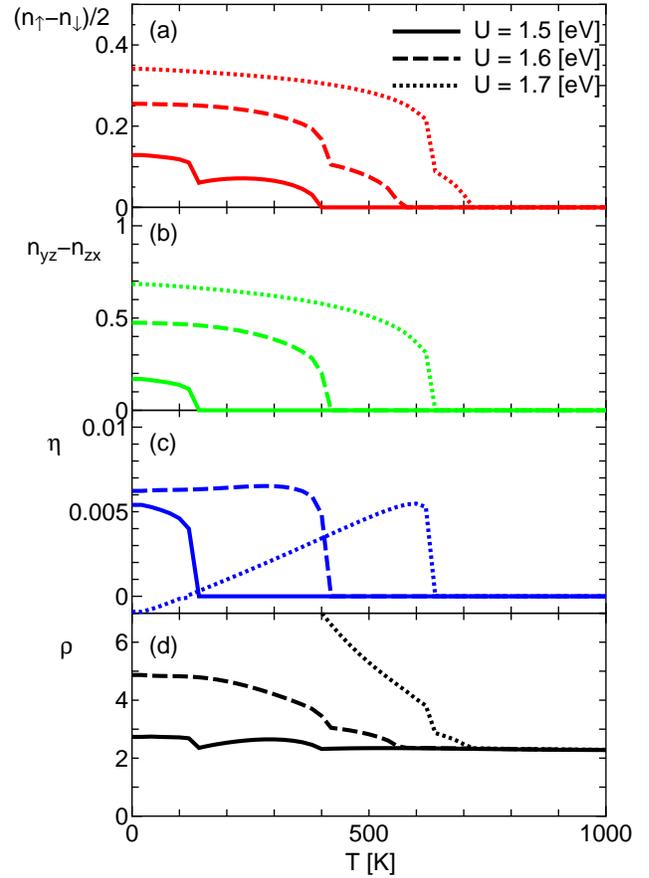} 
\caption{
The temperature dependence of the sublattice magnetization $(n_{\uparrow}-n_{\downarrow})/2$ (a), the orbital order $n_{yz}-n_{zx}$ (b), the charge order $\eta$ (c) and the resistivity $\rho$ (d) for $\Delta \varepsilon_d$ = 1.0 eV at $U = 1.5$ eV (solid lines), $U = 1.6$ eV (dotted lines) and $U = 1.7$ eV (dashed lines). 
}
\label{fig-tdep}
\end{center}
\end{figure}
%%%%%%%%%%%%%%%%%%%%%%%%%%%%%%%%%%%%%%%%%%%%%%%%%%%%%%%%%%%%%%%%%%%%%%%%%%%%

We plot the temperature dependence of the order parameters together with the resistivity for several values of $U$ in Figs. \ref{fig-tdep} (a)-(d). 
%The sublattice magnetization $(n_{\uparrow}-n_{\downarrow})/2$, the orbital order $n_{yz}-n_{zx}$, the charge order $\eta$ and the resistivity $\rho$ as functions of $T$ for $\Delta \varepsilon_{d} = 1.0$ eV are shown in Fig. \ref{fig-tdep} at $U = 1.60$ eV (a), $U = 1.65$ eV (b) and $U = 1.70$ eV (c), respectively. 
We observe the successive phase transitions at $T_{c1}$ and $T_{c2}$ which increase with $U$. 
Below $T_{c1}$, the sublattice magnetization $(n_{\uparrow}-n_{\downarrow})/2$ becomes finite and the AFM is realized as shown in Fig. \ref{fig-tdep} (a). 
Below $T_{c2} ( < T_{c1})$, the orbital order $n_{yz}-n_{zx}$ and the induced tiny charge order $\eta$ become finite and the coexistence state of AFM+CO+OO is realized as shown in Figs. \ref{fig-tdep} (b) and (c). 
The sublattice magnetization is enhanced due to the effect of the orbital order below $T_{c2}$ as shown in Fig. \ref{fig-tdep} (a). 
In the case with large $U = 1.7$ eV, the resistivity $\rho$ shows a small anomaly at $T_{c1}$, while it rapidly increases with decreasing $T$ resulting the insulating grand state as shown in Fig. \ref{fig-tdep} (d). 
On the other hand, in the cases with small $U = 1.5$ and $1.6$ eV, $\rho$ shows kinks at both $T_{c1}$ and $T_{c2}$ resulting the metallic grand state. 
We note that, for larger value of $U > 1.8$ eV, $T_{c1}$ merges with $T_{c2}$ and the direct transition from the paramagnetic state (PM) to the coexistence state of AFM+CO+OO is observed. 

%The resistivity $\rho$ is almost constant while $(n_{\uparrow}-n_{\downarrow})/2$ increases with decreasing $T$. 
%On the other hand, in the AFM+CO+OO state, $n_{\uparrow}-n_{\downarrow}$ and $\rho$ increase with decreasing $T$. 
%and the system becomes insulator when $U$ is large. 
%though it is metal when $U$ is small

%%%%%%%%%%%%%%%%%%%%%%%%%%%%%%%%%%%%%%%%%%%%%%%%%%%%%%%%%%%%%%%%%%%%%%%%%%%%
\begin{figure}[t]
\begin{center}
\includegraphics[scale=0.4]{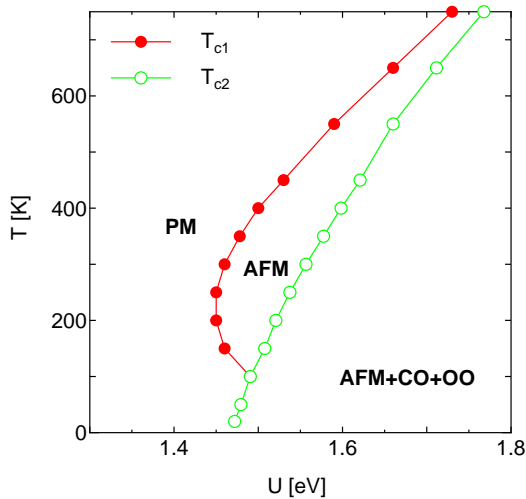}
\caption{
$U$-$T$ phase diagram for $\Delta \varepsilon_{d} = 1.0$ eV. 
%The energy gap increase with the temperature decrease in the AFM+CO+OO state and $\rho (T=0) = \infty$ for $U > 1.7$ eV. 
}
\label{fig-phase}
\end{center}
\end{figure}
%%%%%%%%%%%%%%%%%%%%%%%%%%%%%%%%%%%%%%%%%%%%%%%%%%%%%%%%%%%%%%%%%%%%%%%%%%%%

%%%%%%%%%%%%%%%%%%%%%%%%%%%%%%%%%%%%%%%%%%%%%%%%%%%%%%%%%%%%%%%%%%%%%%%%%%%%
% \begin{figure}[t]
% \begin{center}
% \includegraphics[scale=0.4]{all-tdep_4o83n1uj+1_14_v00e100f10t+2.eps}
% \caption{
% }
% \label{fig-phase}
% \end{center}
% \end{figure}
%%%%%%%%%%%%%%%%%%%%%%%%%%%%%%%%%%%%%%%%%%%%%%%%%%%%%%%%%%%%%%%%%%%%%%%%%%%%

Finally, we show the $U$-$T$ phase diagram including the PM, the AFM and the coexistence state of the AFM+CO+OO for $\Delta \varepsilon_d = 1.0$ eV in Fig. \ref{fig-phase}.
The transition between the PM and the AFM takes place at $T=T_{c1}$, while the transition between the AFM (or the PM) and the AFM+CO+OO takes place at $T=T_{c2}$. 
For $U>1.49$ eV, the system shows the successive phase transitions at $T_{c1}$ and $T_{c2}$ as shown in Fig. \ref{fig-tdep}. 
Remarkably, we also observe the reentrant phase transitions of PM$-$AFM$-$PM for $1.45<U<1.47$ eV, and the reentrant phase transitions of PM$-$AFM$-$PM$-$AFM+CO+OO for $1.47<U<1.49$. 
Such reentrant transitions are due to the schematic structure of the density of states which have a minimum around the Fermi level under the one-dimensional potential $\Delta \varepsilon_d$. 
It is interesting to observe such reentrant transitions in experiments under high pressure. 

%The AFM state is realized only for finite temperature just above the coexistence state of AFM+CO+OO region in the case with $\Delta \varepsilon_d$ = 1.0 eV. 
%When we increase $\Delta \varepsilon_d$, the AFM state is stabilized and finally realized even  for $T = 0$ K. 
%On the other hand, when we decrease $\Delta \varepsilon_d$, the AFM state is suppressed and finally disappears with leaving the coexistence state of AFM+CO+OO (not shown). 
%Therefore, the effect of the one dimensional potential due to the Na order is crucial to realize the successive phase transitions in this system. 

%As mentioned before, the effect of the one dimensional potential $\Delta \varepsilon_d$ stabilize the AFM state. 
%The $T$-$U$ phase diagram with $\Delta \varepsilon_{d} = 1.0$ eV is shown in Fig. \ref{fig-phase}. 
%The AFM transition temperature $T_{c1}$ is larger than the AFM+CO+OO transition temperature $T_{c2}$. 
%This is not trivial because AFM state does not appear when $\Delta \varepsilon_{d}$ is small (not shown). 

%%%%%%%%%%%%%%%%%%%%%%%%%%%%%%%%%%%%%%%%%%%%%%%%%%%%%%%%%%%%%%%%%%%%%%%%%%%%
%\section{Summary and Discussion}
In summary, we investigated the electronic states of the CoO$_2$ plane in the layered cobalt oxides Na$_{0.5}$CoO$_2$ by using the 11 band $d$-$p$ model on the two-dimensional triangular lattice within the Hartree-Fock approximation. 
What we have found are;
(1) The system shows the successive phase transitions where the AFM state is realized below $T_{c1}$ and the coexistence state of AFM+CO+OO is realized below $T_{c2} (< T_{c1})$. 
(2) The effect of the onedimensional potential due to the Na order is crucial for the successive phase transitions. 
(3) The resistivity shows a small anomaly at $T_{c1}$, while it rapidly increases with decreasing $T$ resulting the insulating grand state for large $U$. 
(4) In the coexistence state of AFM+CO+OO, the charge order is induced by the orbital order and is extremely small. 
The obtained results are qualitatively consistent with the experimental results for the successive phase transitions in Na$_{0.5}$CoO$_2$.

In the present paper, we employed the energy bands without the hole pockets. 
In the case with the hole pockets, we also carried out the Hartree-Fock calculations. 
Due to the effect of the one dimensional potential, the resulting Fermi surface is almost the same as that for the case without the hole pockets.
Therefore, the system with the hole pockets also shows the smiler successive phase transitions, although the AFM is slightly stabilized as compared to the case without the hole pockets. 

%Due to the effect of the one dimensional potential, the resulting Fermi surface is almost the same as that for the case without the hole pockets. 
%with finite $\Delta \varepsilon_d$
%The effect of the hole pockets stabilize the AFM state. 

% The small six hole pockets appear in the LDA calculation \cite{Singh} but they do not appear in ARPES experiments \cite{Yang}. 
% Though the calculation result only without the hole pocket was shown in this paper, we also carried out the calculation when there were six hole pockets. 
% The two phase transition is obtained when there were six hole pockets. 
% In these ordered state, the onedimensional Na order is more important than the presence of the hall pockets. 
% However, the increasing of the resistivity $\rho$ according to the phase transition at $T_{c2}$ is a little small. 
% The result without hole pocket seems to be suitable for the experimental result that the resistivity $\rho$ rises below $T_{c2}$. 
%Therefore, it seems to be suitable that there is no hall pocket when comparing it with the experiment result of the rising of resistivity $\rho$ at $T_{c2}$. 

We have proposed a mechanism of the metal-insulator transition at $T_{c2}$, where the orbital order together with a small induced charge order takes place. 
The obtained charge ordering pattern is consistent with the experimentally proposed ordering pattern in Na$_{0.5}$CoO$_2$ \cite{Ning}, although the orbital order, which is crucial in the present theory, was not considered there. 
% is a transition to the coexistence state between antiferromagnetic, charge and orbital ordered states. 
There is another type of the coexistence state of the antiferromagnetism and the orbital order without the charge order, which seems to be consistent with the experiments in K$_{0.5}$CoO$_2$ \cite{Watanabe,Yokoi2} where the charge order has not been observed below $T_{c2}$. 
The explicit calculation including this type of the coexistence state is now under the way.

%Finally, we discuss the possible metal-insulator transition without the charge order at $T_{c2}$, which is expected to realize in K$_{0.5}$CoO$_2$ \cite{Watanabe,Yokoi2}. 

%If this model is correct, very small charge order in the X-direction is caused below $T_{c2}$, and detailed experiments in the future are expected. 
%Moreover, the effect of an electronic correlation was considered, and a quantitative calculation is necessary in the theory. 

%%%%%%%%%%%%%%%%%%%%%%%%%%%%%%%%%%%%%%%%%%%%%%%%%%%%%%%%%%%%%%%%%%%%%%%%%%%%
\section*{Acknowledgments}
The authors thank M. Sato, Y. Kobayashi, M. Yokoi and T. Moyoshi for many useful comments and discussions.
This work was supported by the Grant-in-Aid for Scientific Research from the Ministry of Education, Culture,  Sports, Science and Technology. 

%%%%%%%%%%%%%%%%%%%%%%%%%%%%%%%%%%%%%%%%%%%%%%%%%%%%%%%%%%%%%%%%%%%%%%%%%%%%


\begin{thebibliography}{99}
\bibitem{Takada} 
K. Takada, 
H. Sakurai, 
E. Takayama-Muromachi, 
F. Izumi, 
R. A. Dilanian and 
T. Sasaki, 
Nature {\bf 422} (2003) 53. 
%Superconductivity in twodimensional CoO2 layers

\bibitem{Foo} 
M. L. Foo, 
Y. Wang, 
S. Watauchi, 
H. W. Zandbergen, 
T. He, 
R. J. Cava, and 
N. P. Ong, 
Phys. Rev. Lett. {\bf 92} (2004) 247001. 
%Charge Ordering, Commensurability, and Metallicity in the Phase Diagram of the Layered NaxCoO2

\bibitem{Yokoi} 
M. Yokoi, 
T. Moyoshi, 
Y. Kobayashi, 
M. Soda, 
Y. Yasui, 
M. Sato, and 
K. Kakurai, 
J. Phys. Soc. Jpn. {\bf 74} (2005) 3046. 
%Magnetic Correlation of NaxCoO2 and Successive Phase Transitions of Na0.5CoO2 -NMR and Neutron Diffraction Studies-

\bibitem{Terasaki} 
I. Terasaki, 
Y. Sasago, and 
K. Uchinokura, 
Phys. Rev. B {\bf 56} (1997) R12685. 
%Large thermoelectric power in NaCo2O4 single crystals


\bibitem{Lee} 
M. Lee, 
L. Viciu, 
L. Li, 
Y. Wang, 
M. L. Foo, 
S. Watauchi, 
R. A. Pascal Jr, 
R. J. Cava, and
N. P. Ong, 
Nature Materials {\bf 5} (2006) 537. 

\bibitem{Koshibae}
W. Koshibae, 
K. Tsutsui, and 
S. Maekawa, 
Phys. Rev. B {\bf 62} (2000) 6869.
%Thermopower in cobalt oxides

\bibitem{Motohashi} 
T. Motohashi, 
R. Ueda, 
E. Naujalis, 
T. Tojo, 
I. Terasaki, 
T. Atake, 
M. Karppinen, and 
H. Yamauchi, 
Phys. Rev. B {\bf 67} (2003) 064406. 
%Unconventional magnetic transition and transport behavior in Na0.75CoO2

\bibitem{Bayrakci} 
S. P. Bayrakci, 
I. Mirebeau, 
P. Bourges, 
Y. Sidis, 
M. Enderle, 
J. Mesot, 
D. P. Chen, 
C. T. Lin, and 
B. Keimer, 
Phys. Rev. Lett. {\bf 94} (2005) 157205. 
%Magnetic Ordering and SpinWaves in Na0:82CoO2

\bibitem{Gasparovic} 
Recent Neutron measurements suggest that the $c$-axis oriented moments on the Co$^{3.5-\delta}$ chains are tiny or absent ($<0.04\mu_B$): 
G. Gas$\check{\rm p}$arovi$\acute{\rm c}$, 
R. A. Ott, 
J.-H. Cho, 
F. C. Chou, 
Y. Chu, 
J.W. Lynn, and 
Y. S. Lee,
Phys. Rev. Lett. {\bf 96} (2006) 046403. 
%Neutron Scattering Study of Novel Magnetic Order in Na0:5CoO2

\bibitem{Huang} 
Q. Huang, 
M. L. Foo, 
J. W. Lynn, 
H. W. Zandbergen, 
G. Lawes, 
Y. Wang, 
B. H. Toby, 
A. P. Ramirez, 
N. P. Ong and 
R. J. Cava, 
J. Phys.: Condense. Matter {\bf 16} (2004) 5803. 
%Low temperature phase transitions and crystal structure of Na0.5CoO2

\bibitem{Znadbergen} 
H. W. Zandbergen, 
M. Foo, 
Q. Xu, 
V. Kumar, and 
R. J. Cava
Phys. Rev. B {\bf 70} (2004) 024101. 

\bibitem{Ning} 
F. L. Ning,
S. M. Golin,
A. Ahilan, 
T. Imai, 
G. J. Shu, and 
F. C. Chou, 
Phys. Rev. Lett. {\bf 100} (2008) 086405. 

\bibitem{Watanabe}
H. Watanabe, 
Y. Mori, 
M. Yokoi, 
T. Moyoshi, 
M. Soda, 
Y. Yasui, 
Y. Kobayashi, 
M. Sato, 
N. Igawa, and 
K. Kakurai 
J. Phys. Soc. Jpn. {\bf 75} (2006) 034716. 

\bibitem{Yokoi2} 
M. Yokoi, 
Y. Kobayashi, 
T. Moyoshi, and 
M. Sato, 
to be published in J. Phys. Soc. Jpn. 

\bibitem{Li} 
Z. Li, 
J. Yang, 
J. G. Hou, and 
Q. Zhu, 
Phys. Rev. B {\bf 71} (2005) 024502. 

\bibitem{Indergand} 
M. Indergand,
Y. Yamashita,
H. Kusunose, and 
M. Sigrist, 
Phys. Rev. B {\bf 71} (2005) 214414. 

\bibitem{Bobroff}
J. Bobroff, 
G. Lang, 
H. Alloul, 
N. Blanchard, and 
G. Collin, 
Phys. Rev. Lett. {\bf 96} (2006) 107201. 
%NMR Study of the Magnetic and Metal-Insulator Transitions in Na0:5CoO2: A Nesting Scenario

\bibitem{Choy}
T. P. Choy, 
D. Galanakis, and 
P. Phillips, 
Phys. Rev. B {\bf 75} (2007) 073103. 

\bibitem{Zhou}
S. Zhou and 
Z. Wang, 
Phys. Rev. Lett. {\bf 98} (2007) 226402. 
%Charge and Spin Order on the Triangular Lattice: NaxCoO2 at x 0:5

\bibitem{Yamakawa} 
Y. Yamakawa and 
Y. {\=O}no, 
J. Phys.: Condense. Matter {\bf 19} (2007) 145289. 

\bibitem{Yamakawa2} 
Y. Yamakawa and 
Y. {\=O}no, 
J. Phys. Chem. Solids, in press; arXiv:0806.0911. 

\bibitem{Singh} 
D. J. Singh, 
Phys. Rev. B {\bf 61} (2000) 13397. 
%Electronic structure of NaCo2O4

\bibitem{Yada}
K. Yada and 
H. Kontani
J. Phys. Soc. Jpn. {\bf 74} (2005) 2161.
%Origin of Weak Pseudogap Behaviors in Na0.35CoO2: Absence of Small Hole Pockets

\bibitem{Yada2}
K. Yada and 
H. Kontani
J. Phys. Soc. Jpn. {\bf 75} (2006) 033705.
%Optical Conductivity and Hall Coefficient in High-Tc Superconductors: Significant Role of Current Vertex Corrections

\bibitem{Yang} 
H.-B. Yang, 
Z.-H. Pan, 
A. K. P. Sekharan, 
T. Sato, 
S. Souma, 
T. Takahashi, 
R. Jin, 
B. C. Sales, 
D. Mandrus, 
A. V. Fedorov, 
Z. Wang, and 
H. Ding, 
Phys. Rev. Lett. {\bf 95} (2005) 146401. 

%\bibitem{Mizokawa} 
%T. Mizokawa and 
%A. Fujimori, 
%Phys. Rev. B {\bf 54} (1996) 5368. 
%Electronic structure and orbital ordering in perovskite-type 3d transition-metal oxides studied by Hartree-Fock band-structure calculations

%\bibitem{Balicas}
%L. Balicas, 
%M. Abdel-Jawad, 
%N. E. Hussey, 
%F. C. Chou, and 
%P. A. Lee, 
%Phys. Rev. Lett. {\bf 94} (2005) 236402. 
%ShubnikovovHe Haas Oscillations and the Magnetic-Field-Induced Suppression of the Charge Ordered State in Na0:5CoO2

\end{thebibliography}
\end{document}